\documentclass[twocolumn,showpacs,amsmath,amssymb]{revtex4}

\usepackage{epsfig}

\usepackage{dcolumn}

\usepackage{bm}

\usepackage{amsmath}

\usepackage{amsfonts}

\usepackage{amssymb}

\usepackage{graphicx}%

\newcommand{\beq}{\begin{equation}}
\newcommand{\eeq}{\end{equation}}
\newcommand{\beqn}{\begin{eqnarray}}
\newcommand{\eeqn}{\end{eqnarray}}

\newcommand{\slp}{\raise.15ex\hbox{$/$}\kern-.57em\hbox{$ \partial $}}
\newcommand{\lnA}{\raise.15ex\hbox{$/$}\kern-.57em\hbox{$A$}}
\newcommand{\lnC}{\raise.15ex\hbox{$/$}\kern-.57em\hbox{$C$}}

\newcommand{\A}{\mbox{$\mathcal{A}$}}

\newcommand{\C}{\mbox{$\mathcal{C}$}}

\begin{document}

\title{Quantum pump effect in one-dimensional systems of Dirac fermions}

\author{L. Arrachea}
\affiliation{Departamento de F\'{\i}sica de la Materia Condensada, Universidad
de Zaragoza,  5009 Zaragoza}
\affiliation{BIFI, Universidad de Zaragoza, Corona de Arag\'on 42, 5009 Zaragoza, Spain}

\author{Carlos Na\'on}
\affiliation{Departamento de F\'{\i}sica, Facultad de Ciencias
Exactas, Universidad Nacional de La Plata and IFLP-CONICET, CC 67,
 1900 La Plata, Argentina.}

\author{Mariano Salvay}
\affiliation{Departamento de F\'{\i}sica, Facultad de Ciencias
Exactas, Universidad Nacional de La Plata and IFLP-CONICET, CC 67,
 1900 La Plata, Argentina.}

\date{April 16, 2007}
\begin{abstract}
We investigate the behavior of the directed current in
one-dimensional systems of Dirac fermions driven by local periodic
potentials in the forward as well in backscattering channels. We
treat the problem with
 Keldysh non-equilibrium Green's function
formalism. We present the exact solution for the case of an
infinite wire and show that in this case the dc current vanishes
identically. We also investigate a confined system consistent in
an annular arrangement coupled to a particle reservoir. We present
a perturbative treatment that allows for the analytical
expressions of the dc current in the lowest order of the
amplitudes of the potential. We also present results obtained
from the exact numerical solution of the problem.
\end{abstract}

\maketitle

\section{Introduction}
The possibility of employing ac-fields to induce directed flow of
charge and spin in quantum systems is in the limelight of quantum
transport phenomena. Recent experimental observations of the
quantum pumping effect, including the periodic deformation of the
walls of quantum dots
\cite{SMCG99,GAHMPEUD90,DMH03,VDM05} and the induction of currents
by applying surface acoustic waves in carbon nanotubes \cite{saw}
triggered an important development in the theoretical description
of time-dependent quantum transport
\cite{BTP94,Brouwer98,AA98,ZSA99,AEGS00,
EWAL02,PB01,MCM02,AEGS04,ZLCMcK04,micha1,micha2,micha3,han,lili,lilipr,lilimos}.

The development of new technologies based on quantum phenomena, is
also gauging the search of novel materials to achieve transport in
devices of reduced scale. Among the many structures under
investigation, devices fabricated of graphite are capturing an
increasing attention. From the theoretical side, the Physics of
graphite has the appealing feature of being a realization of a gas
of Dirac fermions in reduced dimensions. In fact, the behavior of
the conductance of single-wall nanotubes \cite{nanotu} seems to be
reasonably explained on the basis of a L\"uttinger liquid,  which
is one of the few models of interacting electrons in condensed
matter Physics that is amenable to be analytically solved. The
Physics of graphene, being a single graphite sheet seems also to
be suitably described as a two-dimensional gas of Dirac fermions
\cite{graphene}. These developments are motivating a renewed
interest in the Physics of Dirac fermions and in the application
of quantum field theoretical methods to shed light on low
dimensional Condensed Matter problems. In particular, a precise
understanding of the effect of time-dependent perturbations on
these systems will be crucial in order to build nanodevices based
on the pumping of charge and spin into a conductor. Recently, we
have solved the problem of an ac-potential in a one-dimension
infinite wire of Dirac fermions, emphasizing the separate role of
the forward and backward channels in the resulting total energy
density and transport properties \cite{carmar}.

\begin{figure}
{\includegraphics[width=80mm,
  keepaspectratio,
clip]{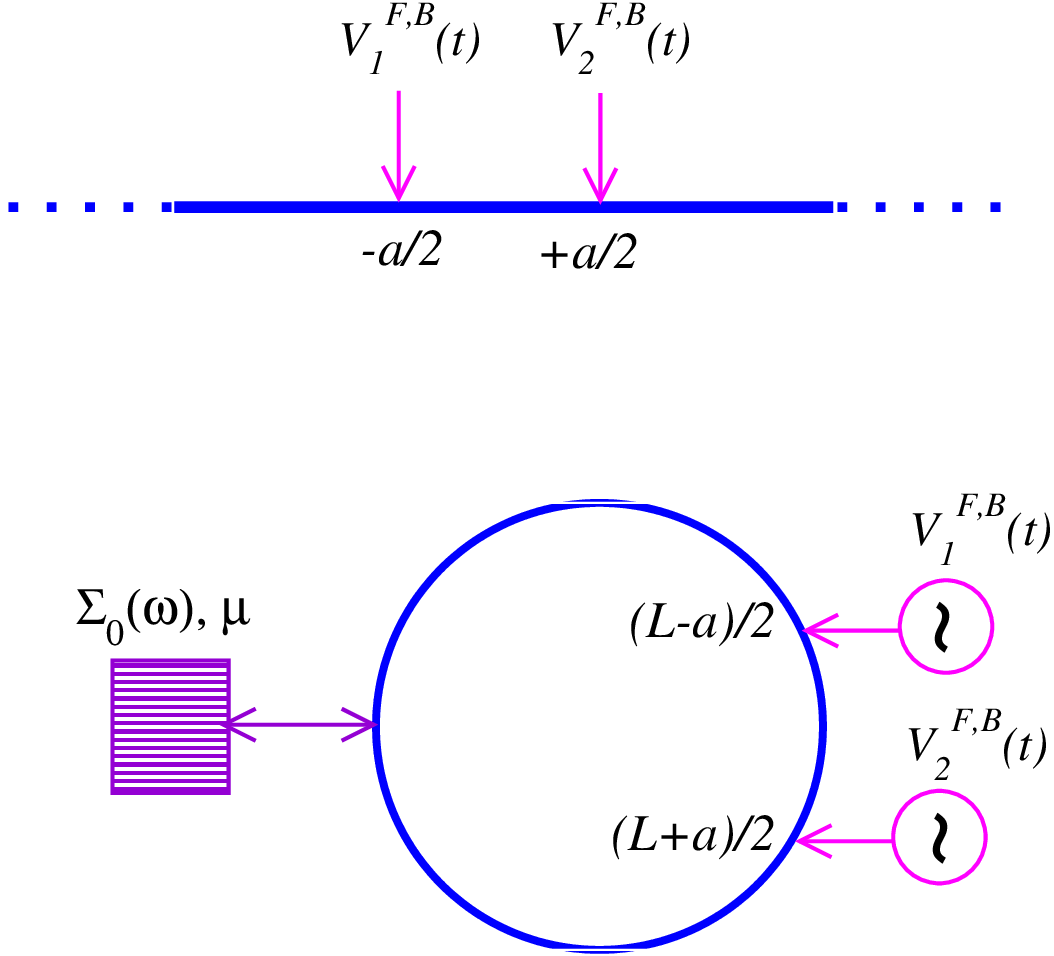}} \caption{(Color online) Sketch of the setup.
Above: Infinite wire driven by periodic fields inducing dynamic
forward and back scattering at the positions $x=\pm a/2$. Bellow:
The ac fields act on a finite wire of length $L$ in anular
geometry. The system is coupled to a particle reservoir a the
position $x=0$. } \label{fig1}
\end{figure}

The breaking of certain symmetries has been singled out as a
necessary condition to generate directed currents. This idea has
been firstly introduced in the context of classical "ratchet"
systems, \cite{rat} and also seems to apply in paradigmatic models
of quantum pumps, like two barrier structures driven by two
harmonic fields oscillating with a phase-lag. \cite{lilipr} We
recall that the pump effect is characterized by the induction of a
directed current with pure ac driving, i.e. without introducing
any explicit dc bias voltage. In this work we adapt the two
oscillating barriers setup to a one-dimensional wire of Dirac
fermions. One of our aims is to identify which are the minimal
ingredients to achieve the quantum pump effect in these systems.
 An infinite wire of Dirac fermions with two local
forward ac fields oscillating with a phase lag can be analytically
solved by generalizing the treatment of Ref. \onlinecite{carmar}
As we will show,  the ensuing induced current has a
vanishing dc component. Then, it is natural to explore the role of
confinement in relation to the appearance of non vanishing dc
currents. This is the main purpose of this work. In particular, we
consider a finite wire
 of length $L$ bent into a ring
and coupled to an external reservoir while driven by  ac
forward and backward fields, as indicated in the sketch of Fig.
\ref{fig1}.
 This setup enables us to analyze the role of
dynamical forward and backscattering in wires containing a
spectrum with many resonant levels, being in contact with a
dissipative environment. We treat the problem with Keldysh
non-equilibrium Green's functions \cite{Keldysh} and find
analytical expressions for the dc current at the lowest order of
perturbation theory in the pumping amplitudes. We also present a
system of exact equations and exact results for the dc current
obtained by the numerical solution of these equations.

The paper is organized as follows. In sections II and III we present the
model and the theoretical treatment. In section IV we present the
exact evaluation of the dc current in an infinite wire with dynamical
fowrd scattering. Section V is devoted to study the transport behavior
of a ring with finite length coupled to a reservoir. 
A summary and
conclusions are presented in
Section VI. Technical details, in particular, the procedure followed to obtain an effective action
through the functional integration of the reservoir's fields, are
described in Appendix A.

\section{Models}
We study a one-dimensional system of  Dirac fermions in two
different setups. Firstly, we consider an infinite wire with two
harmonically time-dependent local potentials applied at the
positions $x_j= {\pm} a/2 $ (see Fig. 1a). This system is
described by an action
\begin{equation}
S = S_{0}+ S_{ac},
\end{equation}
where $S_{0}$ is the linearized free dispersion relation for right
and left-movers (in our unit system $\hbar=v_F=e=1$):
\begin{equation}\label{wire1}
 S_{0} = i\int dx\,dt\,(\psi_r^{\dagger}(t) (\partial_t + \partial_x)
\psi_r(t) +\psi_l^{\dagger}(t) (\partial_t -\partial_x) \psi_l(t)),
\end{equation}
$S_{ac}$ describes forward and backward scattering processes
through the two time-dependent potentials:
\begin{eqnarray}\label{wire2}
S_{ac} & = &
  \sum_{j=1}^{2}\int dx\,dt\, \delta(x- x_j)
\Big[ V^F_j(t)(\psi_r^{\dagger}(t) \psi_r(t)+\nonumber\\
& & +\psi_l^{\dagger}(t) \psi_l(t) )+ V^B_j(t)(\psi_r^{\dagger}(t)
\psi_l(t)+\nonumber\\ & & +\psi_l^{\dagger}(t) \psi_r(t) ) \Big].
\end{eqnarray}

Note that, for simplicity, the dependence of the fields on x has
been omitted in the above equations. To be specific we consider
harmonic potentials with equal amplitudes and a phase-lag
$\delta$: $V^{F,B}_1(t)= V^{F,B}\cos{\Omega_0 t}$ and
$V^{F,B}_2(t)= V^{F,B} \cos{(\Omega_0 t + \delta)}$, where $F, B$
denote forward and backward channels, respectively.

In the second configuration, we assume that the Dirac fermions propagate
 in a ring of
length $L$, which is in contact to a fermionic reservoir. Also in
this case, harmonically time-dependent local potentials are
applied at the positions $x_i= (L {\pm} a)/2 $ . The action
for the full system can be casted in terms of four terms:
\begin{equation}
S = S_{0}+ S_{ac}+ S_{res}+ S_{cont}.
\end{equation}
The reservoir is assumed to be an infinite system of free fermions
which couples to electrons in the ring through a forward
scattering term. The term $S_{cont}$ account for the interaction
of fermions with the reservoir, placed at $x=0$. It reads:
\begin{eqnarray}\label{contact}
 S_{cont} &=& \int dx\,dt\, v_c\,\delta(x)(
\psi_{r}^{\dagger}(t) \chi_{r}(t)  +\psi_{l}^{\dagger}(t)
\chi_{l}(t) +\nonumber\\& &+ \chi_{r}^{\dagger}(t)\psi_{r}(t)
+\chi_{l}^{\dagger}(t) \psi_{l}(t)),
\end{eqnarray}
where the fields $\chi_r^{\dagger}(t), \chi_l^{\dagger}(t)$ denote
degrees of freedom of the reservoir. The latter  can be integrated
out, giving rise to an effective self-energy $\Sigma_0(t-t')$ (see
Appendix A) which leads to the following effective action for the
internal degrees of freedom of the ring:
\begin{eqnarray}\label{effective}
 S_{eff} & = & \int dx\,dt\,dt'\Big( \psi_r^{\dagger}(t) \nonumber \\
& & \Big[ i( \partial_t + \partial_x) \delta(t-t')
-  \delta(x)\,\Sigma_0(t-t') \Big] \psi_r(t') \nonumber \\
& &+ \psi_l^{\dagger}(t) \Big[ i(\partial_t
-\partial_x)\delta(t-t') - \delta(x)\,\Sigma_0(t-t')  \Big]
\psi_l(t')\Big)
\nonumber \\
& & + \sum_{i=1}^{2}\int dx\,dt \,\delta(x- x_i) \, \nonumber \\
& & \times
\Big[ V^F_i(t)(\psi_r^{\dagger}(t) \psi_r(t)+\psi_l^{\dagger}(t)
\psi_l(t) )\nonumber \\
& & + V^B_i(t)(\psi_r^{\dagger}(t) \psi_l(t)+\psi_l^{\dagger}(t)
\psi_r(t) )\Big].
\end{eqnarray}

Our main purpose is to compute the dc component of the time-dependent charge current.
The latter is defined through the condition of the charge conservation:
\begin{equation}
\frac{ d \rho_l(x,t) }{dt} + \frac{ d \rho_r(x,t) }{dt} = \frac{
\partial j(x,t) }{ \partial x}, \label{chacon}
\end{equation}
being $\rho_{\gamma}(x,t)= \langle \psi_{\gamma}^{\dagger}(x,t)
\psi_{\gamma}(x,t) \rangle$, $\gamma=l,r$, which casts:
\begin{equation}
 j(x,t) = \rho_r(x,t)-\rho_l(x,t) . \label{curt}
\end{equation}
The corresponding dc current is
\begin{equation}
 j^{dc}= \frac{1}{\tau_0} \int_0^{\tau_0} dt\,j(x,t), \label{curdc}
\end{equation}
where $\tau_0 = 2 \pi /\Omega_0$ is the period of the
oscillations. Note that, due to the conservation of the charge,
(see eq. (\ref{chacon})), it does not depend on the spacial
coordinate $x$.

\section{Theoretical treatment}
We solve the problem presented in the previous section by recourse
to Keldysh non equilibrium Green's functions. This formalism is
based on the solution of the equation of motion of the matricial
Green's function obtained from the corresponding action.

 From actions (\ref{wire1}) and (\ref{wire2}) it
is easy to show that the matricial Green's function ${\cal G}$ reads:
\begin{eqnarray}\label{Greens}
{\cal G}_{\gamma, \gamma'}(x,x', t, t')  =  \left(\begin{array}{cl} G^{++}_{\gamma, \gamma'} (x,x', t, t') &
G^{+-}_{\gamma, \gamma'} (x,x', t, t')
\\ \\ G^{-+}_{\gamma, \gamma'} (x,x', t, t') & G^{--}_{\gamma, \gamma'}  (x,x', t, t') \end{array} \right),
\end{eqnarray}
where the indexes $+,-$ refer to the closed time path of Keldysh
formalism \cite{Keldysh}, being
\begin{eqnarray}
G^{++}_{\gamma, \gamma'} (x,x', t, t')& = & -i \langle
T\psi_{\gamma}(x,t) \psi^{\dagger}_{\gamma'}(x',t') \rangle,
 \nonumber \\
G^{--}_{\gamma, \gamma'} (x,x', t, t')& = & -i \langle
{\tilde T}\psi_{\gamma}(x,t) \psi^{\dagger}_{\gamma'}(x',t') \rangle,\nonumber \\
G^{+-}_{\gamma, \gamma'} (x,x', t, t')& = & i \langle
\psi^{\dagger}_{\gamma'}(x',t') \psi_{\gamma}(x,t) \rangle,\nonumber \\
G^{-+}_{\gamma, \gamma'} (x,x', t, t')& = & - i \langle
\psi_{\gamma}(x,t) \psi^{\dagger}_{\gamma'}(x',t') \rangle,
\label{ges}
\end{eqnarray}
where $T$ and ${\tilde T}$ denote temporal and anti-temporal
ordering, respectively. These functions satisfy the following
Dyson equation:
\begin{eqnarray} \label{dyson}
& & \Big([i(\partial_t + s_{\gamma} \partial_x)- V^{F}(x,t)]
\delta_{\gamma,\gamma_1} - \nonumber\\& &-V^{B}(x,t)\delta_{\gamma,-\gamma_1}\Big){\cal G}_{\gamma_1,\gamma'}(x,x', t, t') \nonumber \\
& & -\delta(x)\delta_{\gamma,\gamma_1}
 \int dt_1 \Sigma_0(x,t-t_1) {\cal G}_{\gamma_1,\gamma'} (x,x', t_1, t')\nonumber \\
& & = \delta_{\gamma,\gamma'} \delta(x-x') \,\delta_C(t-t')
\end{eqnarray}
where $\delta_C(t-t')$ is the matricial delta function (extended
to the time contour):
\begin{equation}
\delta_{C} (t - t') = \left\{
\begin{array}{ll}\delta(t - t') & ~t , t'~~ \mbox{on}~~ C_+\\
- \delta(t - t') & ~ t , t'~~ \mbox{on}~~ C_-\\ 0 &
~\mbox{otherwise}
\end{array} \right.
\end{equation}
$V^{F,B}(x,t)=\delta(x-x_1)V_1^{F,B}(t)+\delta(x-x_2)V_2^{F,B}(t)$,
$s_{\gamma}=+,-$ for the right and left movers, respectively, and
\begin{eqnarray}\label{self}
\Sigma_0(t-t') =  \left(\begin{array}{cl} \Sigma_0^{++}(t-t') &
 \Sigma_0^{+-}(t-t')
\\ \\  \Sigma_0^{-+}(t-t') &
\Sigma_0^{--} (t-t') \end{array} \right).
\end{eqnarray}
In terms of Green's functions, the current (\ref{curt}) reads
\begin{equation}\label{curtg}
 j(x,t)= -i[G^{+-}_{rr}(x,x,t,t)-G^{+-}_{ll}(x,x,t,t)].
\end{equation}

As shown in the next section,
 for the infinite system with $\Sigma_0(t-t')=0$,
 the Green's function admits a factorization in terms of
the noninteracting (equilibrium) Green's function, at least for ac
potentials that behave as forward scatterers. One then obtains an
analytical expression for $j(x,t)$ with vanishing time average.
This result can be extended perturbatively to the backscattering
case.

The situation changes dramatically when fermions propagate in a
ring connected to a reservoir. As a consequence of a non-instantaneous
term associated to the coupling to the reservoir, the equation of motion
for the
Green's function ceases to be exactly solvable even for the
forward channel and its solution must be found by recourse to perturbation
theory or numerical methods.

\section{Infinite wire}
For the infinite wire, we can adapt the treatment of Ref.
\cite{carmar}, to the case of more than one oscillating
potentials. For the case of pure forward ac potentials (i.e.
$V^B_i=0, \forall i$), $j(x,t)$ can be exactly evaluated and the
result is a purely ac current. To show this we solve the Dyson
equation (\ref{dyson}) for $\Sigma_0(t-t')=0$ and two oscillating
barriers placed at $x=\pm a/2$. Specializing eq. (\ref{dyson}) to
the case $V^B=0$ and considering first, for illustrative purposes
the Green's function for right-movers, results
\begin{equation}\label{dywi}
(i(\partial_t + \partial_x)- V^F(x,t)){\cal G} (x,x', t, t')=
\delta(x-x')\,\delta_C(t-t')
\end{equation}
It is very useful to note that the solution of the above
differential equation for the matrix ${\cal G}$ can be obtained in
terms of the free function ${\cal G}^{0}$, the latter being
the solution of (\ref{dywi}), with $V^F(x,t)=0$. The result is:
\begin{equation}\label{ap1}
{\cal G}(x,t; x', t') = {\cal G}^0(x - x', t - t') \exp [ \beta
(x, t) - \beta (x', t')],
\end{equation}
with
\begin{equation}
\beta(x,t) = \int dx'\,dt'\,G^{0,R}(x - x', t - t')V^F(x',t'),
\end{equation}
where $G^R(x - x', t - t')=G^{0,++}-G^{0,+-}$ is the equilibrium retarded Green's function
of the free system. The above equation can be readily solved to
obtain:
\begin{eqnarray}
\beta(x,t) = -iV^{F}\Big( \theta(x-a/2)\,\cos{\Omega_0(
t-(x-a/2))}+ \nonumber \\+ \theta(x+a/2)\,\cos{(\Omega_0
(t-(x+a/2)) +\delta)} \Big)
\end{eqnarray}
 Having now $\beta(x,t)$ we have, via
equation (\ref{ap1}), every component of ${\cal G}$. Of course,
the same steps can be followed to evaluate the Green's functions
for left-movers. The functions $G^{+-}_{\gamma,\gamma}(x,x',t,t')$ can be used to
compute the current from eq. (\ref{curtg}). This leads to
\begin{eqnarray}\label{jwire}
 j(x,t)&=&\frac{\Omega_0\,V^F}{2\pi} \Big(\theta(x+a/2) \sin[\Omega_0(t-(x+a/2))]
  \nonumber \\
 &+&\theta(x-a/2) \sin[\Omega_0(t-(x-a/2))+\delta]\nonumber \\
 &+& \theta(-(x+a/2))
 \sin[\Omega_0(t+x+a/2)] \nonumber \\
 &+&\theta(-(x-a/2))
 \sin[\Omega_0(t+x-a/2)+\delta]\Big)
\end{eqnarray}
which yields $j_{wire}^{dc}=0$.

In the case $V^B \neq 0$ it is not possible to arrive at a closed
expression for the current, one is forced to perform a
perturbative expansion. We have computed $j(x,t)$ up to second
order in $V^B$, showing that even for this case the time average
of $j(x,t)$ vanishes for the infinite wire.

\section{Finite ring in contact to a reservoir}
\subsection{Evaluation of the Green's function}
In the case of the annular setup, the presence of a self-energy
 in the effective
action $S_{eff}$ renders the procedure of the previous section unsuitable
even in the case of pure forward dynamical scattering. However,
the Green's functions can be formally exactly calculated by
following a similar route to the one adopted in the solution of
time-dependent lattice Hamiltonians \cite{lili}. In what
follows we summarize this procedure. It is  convenient to combine
the different components of the Green´s function matrix to define
a retarded Green's function:
\begin{eqnarray}
G^R_{\gamma, \gamma'} (x,x', t, t') & = & \Theta(t-t')
[G^{-+}_{\gamma, \gamma'} (x,x', t, t') \nonumber \\
& & -G^{+-}_{\gamma, \gamma'} (x,x', t, t'),
\label{gr}
\end{eqnarray}
and to perform a Fourier transform with respect to the difference of times:
\begin{equation}
G^R_{\gamma, \gamma'} (x,x', t, t') = \int \frac{d \omega}{2
\pi}\, G^R_{\gamma, \gamma'} (x,x', t, \omega )\, e^{- i \omega
(t- t')}. \label{four}
\end{equation}
Substituting it in the integral representation of  the Dyson's
equations (\ref{dyson}), the following set of linear equations is
obtained \cite{lili}:
\begin{eqnarray}
& &G^R_{\gamma, \gamma'}(x,x', t, \omega) = G^0_{\gamma} (x,x',
\omega )
\nonumber \\
 & &
+ \sum_{j=1}^2 e^{-i( \Omega_0 t - \delta_j) }
G^R_{\gamma,\gamma' } (x,x_j, t, \omega + \Omega_0) \nonumber \\
 & & \times \Big[ V^F  \delta_{ \gamma, \gamma'} +
 V^B \delta_{\overline{\gamma}, \gamma'}
  \Big] G^0_{\gamma'} (x_j,x', \omega)
\nonumber \\
 & & + \sum_{j=1}^2 e^{(i \Omega_0 t + \delta_j)}
G^R_{\gamma, \gamma'} (x, x_j, t, \omega - \Omega_0) \nonumber \\
 & & \times \Big[ V^F  \delta_{ \gamma, \gamma'} +
 V^B \delta_{\overline{\gamma}, \gamma'}
  \Big] G^0_{ \gamma'} (x_j,x', \omega) ,
\label{dyret}
\end{eqnarray}
being $\overline{r}= l$ and $\overline{l}= r$.
The function  $G^0_{\gamma} (x_j, x', \omega)$ is the retarded Green's
function  of Dirac left and right movers of the ring in contact to
the reservoir with $V_i(t)=0$. This function can be calculated
by considering the corresponding Dyson equation, which in the
present case takes the following simple form:
\begin{eqnarray}
G^0_{ \gamma}(x,x',\omega) & = &g_{\gamma}^{0}(x,x',\omega)\nonumber \\
 & &+G^0_{ \gamma}(x,0,\omega)
\Sigma_0^R(\omega)g^0_{ \gamma}(0,x',\omega), \label{g0}
\end{eqnarray}
where  $g_{\gamma}^{0}(x,x',\omega)$ are the free retarded
functions, corresponding to the ring without driving and uncoupled from
the reservoir, while $\Sigma_0^R(\omega)$ is the Fourier transform
of the retarded self-energy: $\Sigma_0^R(t-t')=\Theta(t-t')
[\Sigma_0^{-+}(t - t') -\Sigma_0^{+-}(t - t')]$.
 It can be in general expressed in terms of its spectral
representation $\Gamma(\omega)= - 2\mbox{Im}[\Sigma_0^R(\omega)]$
as follows:
\begin{equation}\label{selfres}
\Sigma_0^R(\omega) = \int_{-\infty}^{\infty} \frac{d \omega´}{2 \pi} \frac{\Gamma(\omega)}{\omega- \omega´+ i 0^+ }.
\end{equation}

The equation for the lesser function reads:
\begin{eqnarray}
G^{+-}_{\gamma, \gamma'} (x,x', t, t') &=& \sum_{\gamma_1=l,r}
 \int_{-\infty}^{\infty} \frac{d\omega}{2 \pi} e^{-i \omega(t-t')}
G^R_{\gamma, \gamma_1} (x,0, t, \omega ) \nonumber \\
 & & \times
\Sigma_0^{+-}(\omega)
[G^R_{\gamma', \gamma_1} (0,x', t', \omega )]^*. \label{dyles}
\end{eqnarray}
This function can be  used to evaluate current from (\ref{curtg}).
Recalling that $\Sigma_0^{+-}(\omega)= i f(\omega)
\Gamma(\omega)$, where $f(\omega)= \theta(-(\omega - \mu))$ is the
Fermi function (we shall consider the zero-temperature case, $\mu$
is the chemical potential of the reservoir), the dc current can, thus, be
expressed as follows:
\begin{equation}\label{current}
j^{dc}= \int \frac{d\omega}{2\pi}f(\omega)T(\omega),
\end{equation}
where the functions $T(\omega)$ collects the spectral
contributions to the current from the forward and backward
channels. Concretely:
\begin{eqnarray}\label{tf}
T(\omega) & = & \frac{1}{\tau_0}\int_{0}^{\tau_0} dt
\Big({\mid G^R_{rr}(x,0,t,\omega)\mid}^2 \nonumber \\
 & &
-{\mid G_{ll}^R(x,0,t,\omega)\mid}^2 + {\mid G^R_{rl}(x,0,t,\omega)\mid}^2 \nonumber \\
 & &
-{\mid G_{lr}^R(x,0,t,\omega)\mid}^2 \Big) \Gamma(\omega),
\end{eqnarray}

Therefore, the evaluation of the retarded Green´s functions leads
us to the current. To this end, we must solve Dyson's equation
(\ref{dyret}).

\subsection{Perturbative expansion in the pumping amplitudes}
In view of the obvious difficulties of solving equation
(\ref{dyret}), we propose in this section a perturbative solution
that allows us to evaluate $j^{dc}$ to order ${\cal O}(|V^F|^2,
|V^B|^2)$. We will obtain an analytical expression for $j^{dc}$,
through the ``transmission function'' $T(\omega)$, valid for any kind of
fermionic reservoir interacting via forward scattering with
the fermions in the ring, like the one described by $S_{cont}$ introduced
in eq. (\ref{contact}).

Before going to the perturbative treatment, let us note that given the
Green's function of the free system:
\begin{equation}
g_{r}^{0}(x,x',\omega)=g_{l}^{0}(x',x,\omega)= \lim_{\Lambda \rightarrow \infty}
\frac{\pi}{\Lambda} \frac{\exp{i \omega
(x-x'-L/2)}}{2\sin{\omega L/2}}, \label{glib}
\end{equation}
Eq. (\ref{g0}) can be exactly solved:
\begin{equation}
G^0_{r}(x,x',\omega)=G^0_{l}(x',x,\omega)=\lim_{\Lambda \rightarrow \infty}
\frac{\pi}{\Lambda} \frac{\exp{i \omega
(x-x'-L/2)}}{h(\omega)},\label{G0}
\end{equation}
with
\begin{equation}
h(\omega)= 2\sin{\omega L/2}-(\Sigma_0^R(\omega))e^{-i\omega L/2}. \label{h}
\end{equation}
The normalization factor ${\pi}/{\Lambda}$ is introduced because
the functions $G^0_{\gamma}(x,x´,\omega)$ and
$g_{\gamma}^{0}(x,x',\omega)$ are not regular, in the sense that
the densities of states $\rho_{\gamma}(x,x,\omega)=-2 \mbox{Im}[
G^0_{\gamma }(x,x´,\omega)]$ and $\rho^0_{\gamma}(x,x,\omega)=-2
\mbox{Im}[ g^0_{\gamma }(x,x´,\omega)]$  do not integrate to $1$.
Actually, the ensuing  integrals diverge and, for this reason,
upper and lower cutoffs $\pm \Lambda$ are introduced to ensure the correct sum rule.
In the case of the Green's function for the ring coupled to the reservoir,
the normalization is the same as for the free system provided that
the reservoir has  a finite bandwidth $W$, such that
 $\Gamma(w) =0$ for $|\omega|>W$, and $\Lambda >> W$.

 Given a concrete model for the reservoir, which
implies assuming a concrete functional form for the spectral
density $\Gamma(\omega)$, one has a definite function $h(\omega)$
(See (\ref{h})). This, in turn, leads to the retarded functions
for the system without time-dependent potentials, given in
(\ref{g0}). These functions are our building blocks for the
perturbative solution of the full retarded Green´s function
$G_{\gamma \gamma }^R(x,x',t,\omega)$. Solving equation
(\ref{dyret}) up to second order in the couplings and replacing
the results in (\ref{tf}), we get
\begin{eqnarray} \label{tfp}
T(\omega)& = & \frac{-\pi^2}{\Lambda^2}
\frac{\Gamma(\omega)}{{\mid
h(\omega)\mid^2}}\sin{\delta}\,\times \nonumber \\
& & \Big(V_{F}^2\,\sin{\Omega_0 a}\,
[T^{(1)}(\omega)+T^{(2)}(\omega)]- \nonumber\\ & & -V_B^2
[T^{(3)}(\omega)+T^{(4)}(\omega)]\Big) ,
\end{eqnarray}

\begin{figure}
{\includegraphics[width=80mm,
  keepaspectratio,
clip]{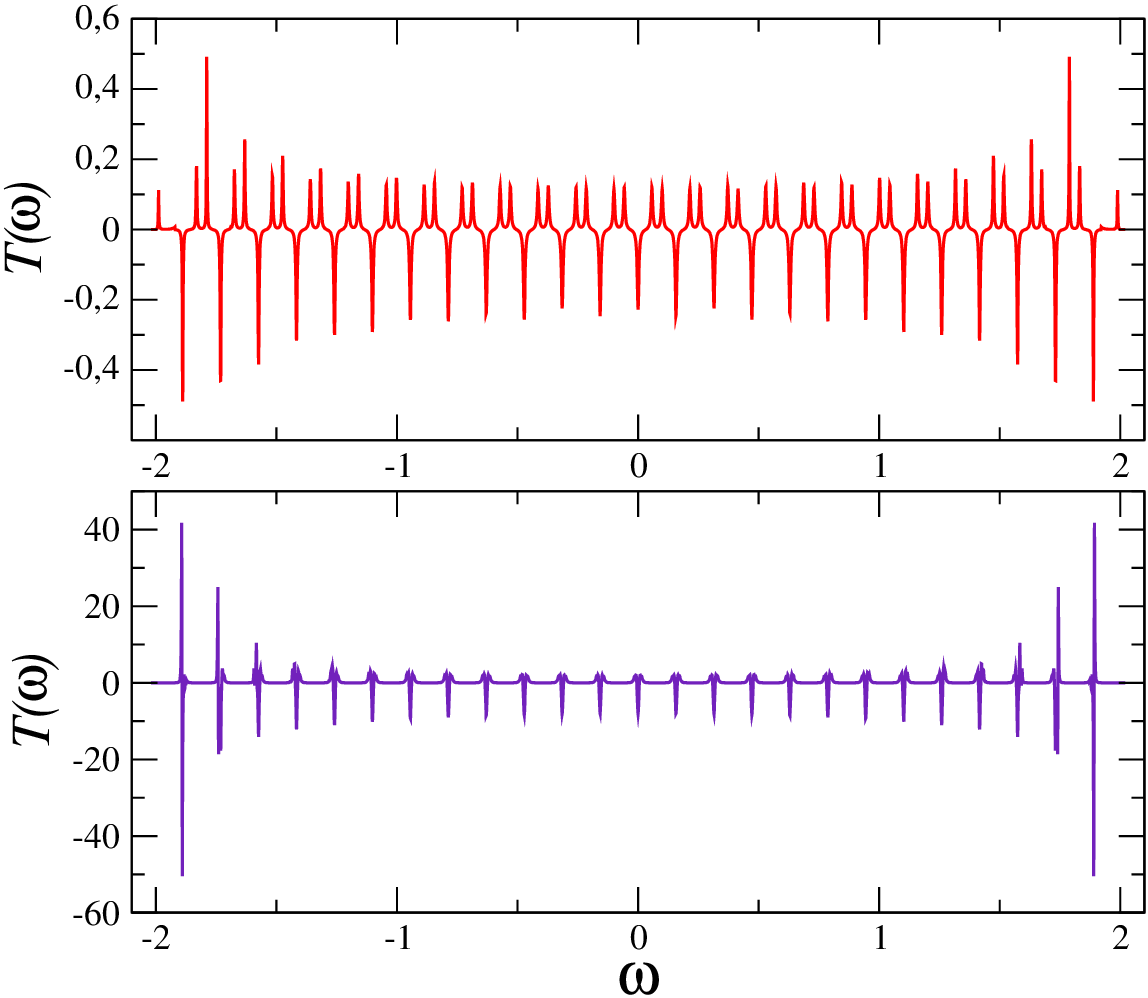}} \caption{(Color online) The function $T(\omega)$
for pure forward scattering. Parameters are: $V_F=0.2$,
$\delta=\pi/2$, $a=2$, $L=40$. We consider a reservoir with
bandwidth $W=16$ and a coupling constant $v_c=0.1$. Upper and
lower panels
 correspond
to $\Omega_0=0.1, 0.15$, respectively.}
\label{fig2}
\end{figure}

where
\begin{equation}\label{t1}
T^{(1)}(\omega)=  \frac{1}{\mid
h(\omega+\Omega_0)\mid^2}+\frac{1}{\mid h(\omega-\Omega_0)\mid^2},
\end{equation}
\begin{equation}\label{t2}
T^{(2)}(\omega)= 2  Re \frac{e^{i\omega
L}}{h^*(\omega)}\left(\frac{e^{i\Omega_0
L/2}}{h^*(\omega+\Omega_0)}+\frac{e^{-i\Omega_0
L/2}}{h^*(\omega-\Omega_0)}\right),
\end{equation}
\begin{multline}
T^{(3)}(\omega)= 2  Re \frac{e^{i\omega
L}}{h^*(\omega)}\Big(\frac{\sin{((2\omega-\Omega_0)
a)}e^{i\Omega_0 L/2}}{h^*(\omega+\Omega_0)}- \nonumber
\\ \frac{\sin{((2\omega+\Omega_0) a)}e^{-i\Omega_0
L/2}}{h^*(\omega-\Omega_0)}\Big)
\end{multline}
\begin{equation}
T^{(4)}(\omega)= \frac{\sin{(2\omega-\Omega_0) a}}{\mid
h(\omega-\Omega_0)\mid^2}-\frac{\sin{(2\omega+\Omega_0) a}}{\mid
h(\omega+\Omega_0)\mid^2}.
\end{equation}
In equation (\ref{tfp}) we have omitted, for sake of simplicity,
the limit $\Lambda \rightarrow \infty$. Let us stress again that
these formulas do not depend on the specific choice of the
function $\Gamma(\omega)$. This allows us to obtain some general
properties of quantum pumping in the annular geometry which are
independent of the details of the reservoir. As a first point, we
stress that unlike the case of the infinite wire, in the finite
ring, forward scattering is enough to generate a directed current.
Second, it is interesting to note that we recover the behavior
$j^{dc} \propto \sin(\delta)$ predicted in Ref. \cite{Brouwer98}
for a two-barrier quantum pump in a linear setup, also obtained
within Floquet scattering matrix formalism \cite{micha1} as well
as within Green's function formalism in tight-binding models
\cite{lili,lilipr}. As a third point, for $\delta \neq mod(\pi)$
and $\Omega_0 a\ll1$, the dc current behaves like $j^{dc} \propto
\Omega_0 $, which is the typical behavior of the so called
``adiabatic'' regime of quantum pumps \cite{Brouwer98,EWAL02,micha1,micha2,micha3}. A final and
important point to note is the role of geometrical factors, like
the length $L$ and the situation of the barriers $a$. In
particular, the prefactor in (\ref{tfp}) indicates that for weak
amplitudes and special combinations of parameters such that
$\Omega_0 a=\pi$, the dc current is inhibited in the forward
channel. In addition, geometrical factors lead to sign modulation
of the induced dc current as a function of the chemical potential
$\mu$, as we will discuss in more detail in the next section.

\subsection{Exact numerical solution}
\begin{figure}
{\includegraphics[width=80mm,
  keepaspectratio,
clip]{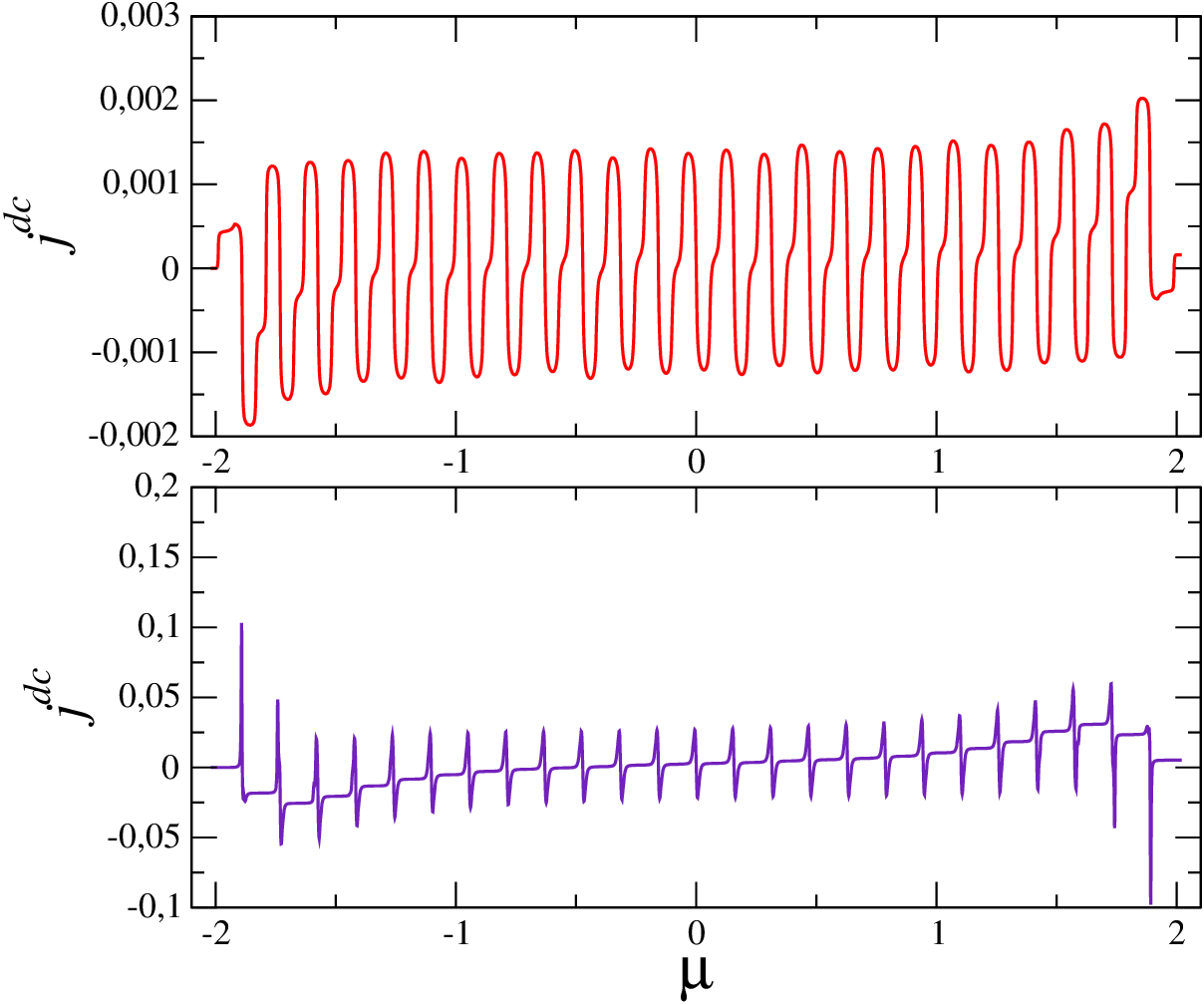}}
\caption{(Color online)
dc current $J^{dc}$ as a function of the
chemical potential of the reservoir for pure forward scattering.
Parameters are the same as in Fig. \ref{fig2}.}
\label{fig3}
\end{figure}
In this section we present results obtained from the numerical
exact solution of the Dyson's equation (\ref{dyret}). In order to
avoid the problem of the regularization of the Green's function
discussed in the previous section,
we consider in this section functions
\begin{equation}
g^0_{\gamma} (x,x',\omega) = \frac{1}{2 N+1}
\sum_{n=-N}^{N} \frac{e^{-i k_n(x-x') }}{\omega \pm k_n + i 0^+},
\end{equation}
being $k_n= 2\pi/L$ while the upper (lower) sign corresponds to
$\gamma= r,l$. This function coincides with (\ref{glib}) in the
limit of $N \rightarrow \infty$, and has well defined spectral
properties, i.e. it satisfies the usual sum rule for the spectral
density. The drawback of this Green's function is that Dyson's
equation (\ref{g0}) cannot be analytically solved. In any case,
the equilibrium Green's functions can be numerically evaluated
from the expression:
\begin{eqnarray}
G^0_{\gamma} (x,x',\omega) & = & g^0_{\gamma} (x,x',\omega) \nonumber \\
& & + \frac{g^0_{\gamma} (x,0,\omega) g^0_{\gamma} (0,x',\omega)}
{1- \Sigma_0^R(\omega) g^0_{\gamma} (0,0,\omega)},
\end{eqnarray}
and used as input to the numerical evaluation of (\ref{dyret}).

So far, we have not introduced any specific model for the
reservoir. In the numerical procedure we have represented the
latter system in terms of a semicircular density of states of
bandwidth $W$ and a coupling $v_c$ between ring and reservoir. The
ensuing self-energy is defined from (\ref{selfres}) with
$\Gamma(\omega)= 4 |v_c|^2 \Theta(W-|\omega|)
\sqrt{W^2-{\omega}^2} /W^2 $.

\begin{figure}
{\includegraphics[width=80mm,
  keepaspectratio,clip]{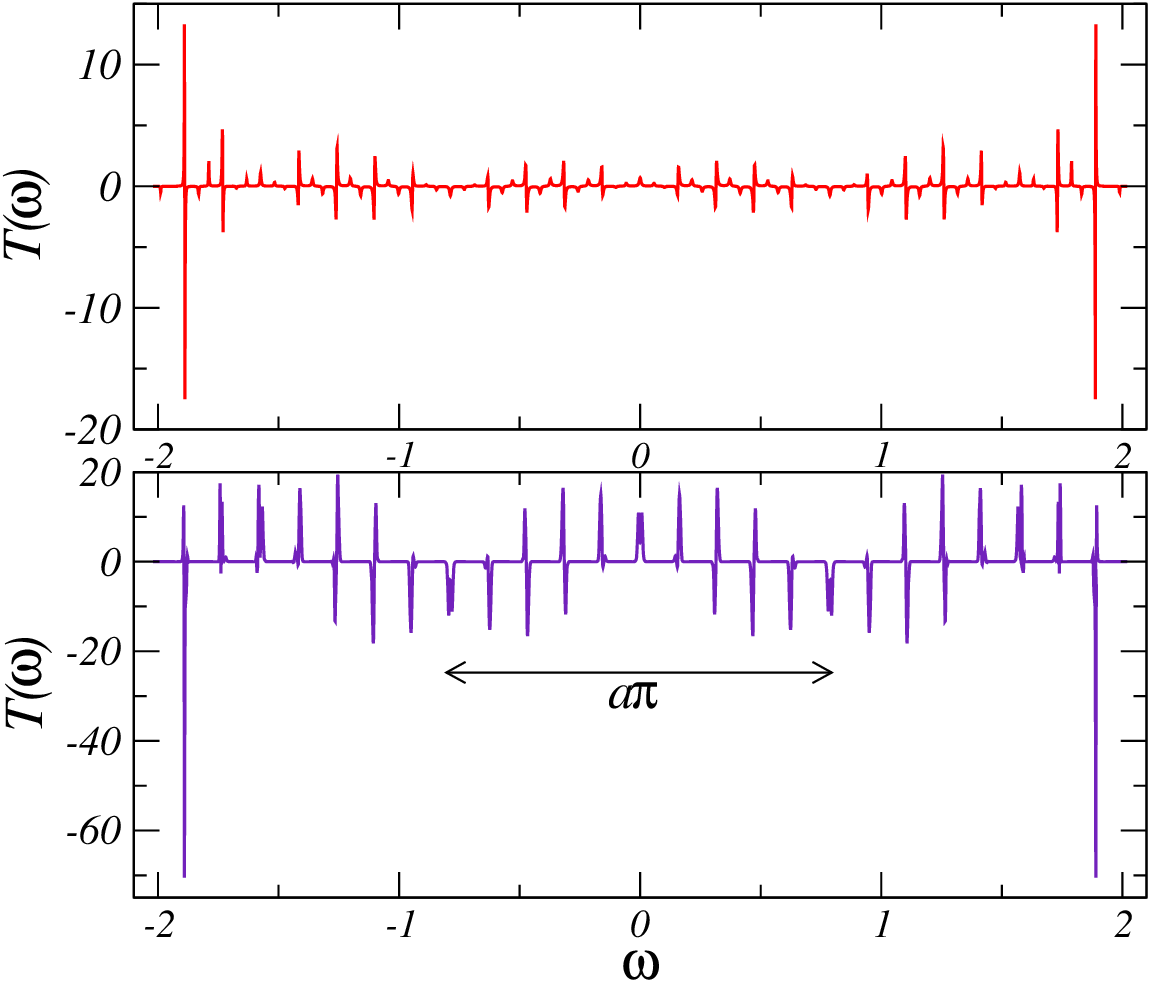}}
\caption{(Color online) The function $T(\omega)$ for pure backward
scattering with $V_B=0.2$. Other details are the same as in Fig.
\ref{fig3}.} \label{fig4}
\end{figure}

Results for the function $T(\omega)$ in the case of pure forward
scattering are shown in Fig. \ref{fig2}. This function displays a
landscape of peaks and anti-peaks that are in accordance with the
structure predicted by the perturbative solution. For $V_B=0$ only
the functions $T^{(1)}(\omega)$ and $T^{(2)}(\omega)$ contribute
(see Eqs. (\ref{tfp}), (\ref{t1}) and (\ref{t2})). Notice that
these functions have peaks centered at the zeros of the function
$h(\omega \pm \Omega_0)$ with $\Sigma_0(\omega)=0$, which
correspond to the positions of the energy levels of the uncoupled
ring, i.e. at $\omega \pm \Omega_0= 2 n \pi/L$, with $n$ integer.
The modulation with period $\omega L$ of $T^{(2)}(\omega)$ also
explains the pattern of alternative maxima and minima of
$T(\omega)$ at
 the energy levels of the uncoupled ring. In the lower panel of Fig.
\ref{fig2} it is shown the behavior of $T(\omega)$ for $\Omega_0
\sim 2 \pi /L$, i.e. close to resonance with the characteristic
frequency associated to the level spacing of the uncoupled ring.

  At zero temperature, the integral of $T(\omega)$ over energies
below the chemical potential $\mu$ gives the dc component of
current for a given $\mu$. The behavior of $J^{dc}$ as a function
$\mu$ is shown in Fig. \ref{fig3} for a driven system with the
same parameters of \ref{fig2}. In the upper panel, corresponding
to $\Omega_0 < 2 \pi/L$,
 a sequence of plateaus of width $\Omega_0$ are clearly distinguished.
In addition sign inversion of the current takes place at points
distant in the level spacing $2 \pi/L$. In the case of the lower panel
(close to resonance) the plateaus approximately coincide with the
level spacing ($\sim \Omega_0$).

The function $T(\omega)$ for pure dynamical backscattering
($V_F=0$) is shown in Fig. \ref{fig4}. The behavior of this
function is also in agreement with the one predicted by the
perturbative solution. In this case, only the functions
$T^{(3)}(\omega)$ and $T^{(4)}(\omega)$ contribute. As in the case
of pure forward scattering, these functions have a sequence of
peaks and anti-peaks defined from $\omega \pm \Omega_0 = 2 \pi
/L$. However in the present case, the structure is richer than in
the case of forward scattering. This is due to the existence of
two modulating factors: $e^{i \omega L}$ in $T^{(3)}(\omega)$, and
the $\sin[(2 \omega \pm  \Omega_0) a]$ in  $T^{(3)}(\omega)$ as
well as in $T^{(4)}(\omega)$. While the first factor introduces a
periodicity which is coincident with the level spacing, the second
one introduces a periodicity $\pi a$ that is defined by the
position at which the ac potentials are applied. In addition, for
backscattering,
 different terms contribute with different
signs, while in the case of forward scattering all the contributions
are additive.

\begin{figure}
{\includegraphics[width=80mm,
  keepaspectratio,clip]{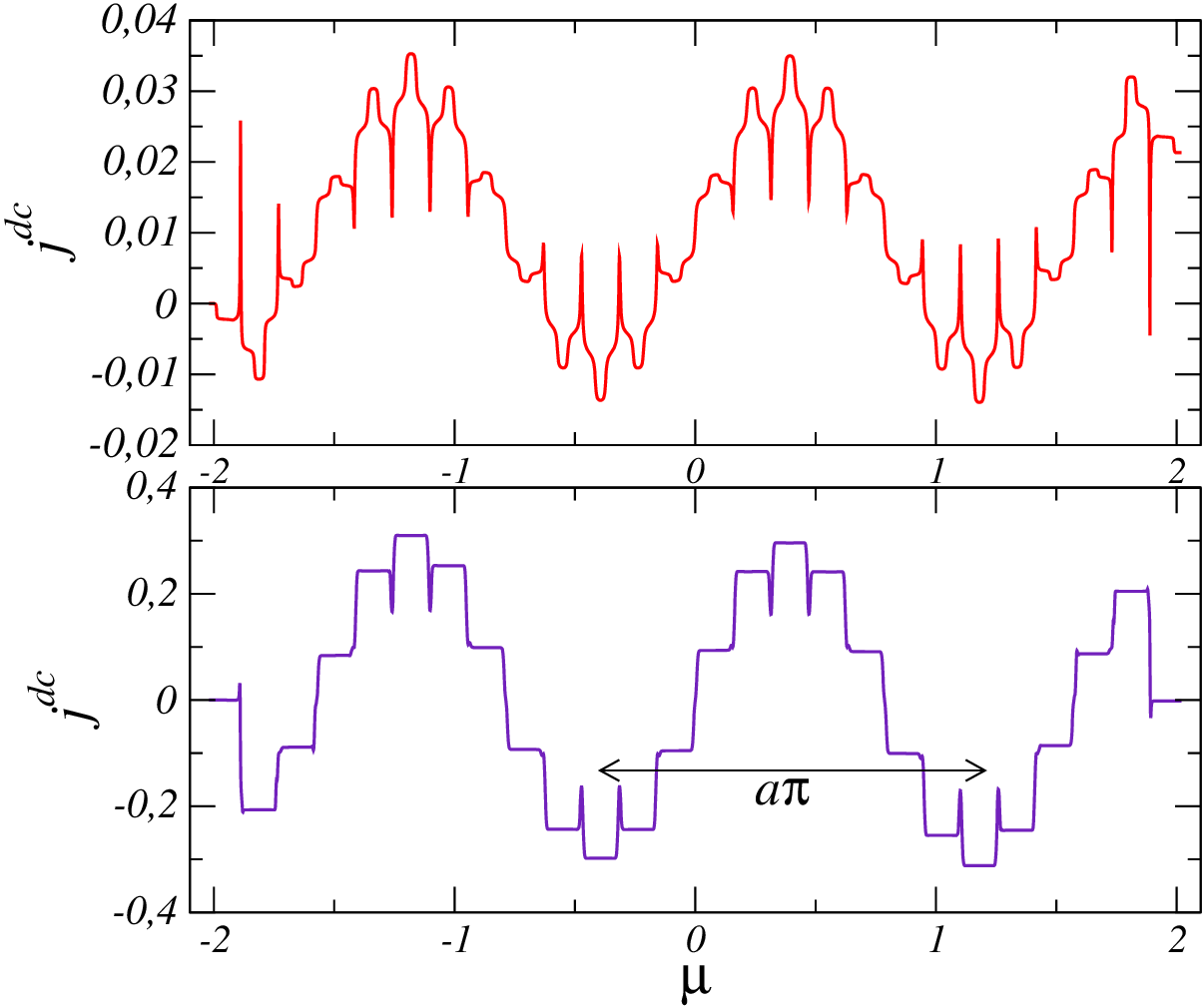}}
\caption{(Color online)
 dc current $J^{dc}$ as a function of the
chemical potential of the reservoir for pure  backward scattering.
Other details are the same as in Fig. \ref{fig4}.}
\label{fig5}
\end{figure}

In Fig. \ref{fig4}, an envelope with period  $a \pi$ can be
clearly distinguished and it is indicated with an arrow in the
lower panel. The corresponding dc current is shown in Fig.
\ref{fig5}. For low pumping frequencies (upper panel) as well as
for frequencies close to resonance (lower panel),  the behavior of
the directed current as a function of the chemical potential
displays sign inversions and a sequence of maxima and minima
separated in  $\pi a/2$. In addition, there is a fine structure of
plateaus of width $\Omega_0$, which is more pronounced in the case
of low $\Omega_0$ (upper panel). Also notice, and additional
structure related to the level spacing energy $2 \pi/L$ of the
ring.

Thus, dynamical forward and backscattering mechanisms leave
clearly different fingerprints in the behavior of the induced
directed current as a function of the chemical potential of the
reservoir. While both processes are sensitive and give rise to
patterns related to the level spacing and the pumping frequency,
backscattering leads to a response that also provides information
on the positions at which the pumping potentials are applied. This
feature accounts for interference processes in the electronic
propagation that take place when electrons emit or absorb quanta
and invert at the same time their direction of motion at the
backscattering centers.

\section{Summary and conclusions}
In this work we have studied quantum pumping induced by two local
ac potentials in the forward and backward channels, oscillating
with a single frequency and a phase lag, in a one-dimensional
system of Dirac fermions. This problem can be exactly solved
analytically in the case of an infinite wire  and pure forward
scattering. The exact solution shows that the charge current in
this case is purely ac without a dc component.
 For this reason, we have investigated a
confined system of annular shape in contact to a particle
reservoir. We have employed a formalism based on Keldysh
non-equilibrium Green's functions to solve this problem. In this
case, an analytical solution is available a the lowest order in
the pumping amplitudes, while for arbitrary amplitudes, it can be
exactly solved by numerically evaluating a Dyson equation for the
retarded Green's functions.

We have shown that a directed current establishes in the confined
geometry with dynamical forward as well as backscattering.
As a function of the phase-lag $\delta$, the current  shows the $
\propto \sin \delta$ behavior already presented in the literature
in the framework of adiabatic descriptions \cite{Brouwer98},
models of plane-wave \cite{micha1} and tight-binding electrons,
\cite{lili,lilipr} and experimentally observed in Ref.
\onlinecite{SMCG99}. Interestingly, in the case of pure forward
scattering there is an additional factor $ \propto \sin( \Omega_0
a) $, denoting interference effects related to the geometrical
arrangement. As a function of the chemical potential of the reservoir, forward
and backward scattering display features in the behavior
of the dc current which are related to the wire level spacing
the pumping frequency.  In the case of backscattering, there is
additional structure related to geometrical parameters.

\section{Acknowledgments}
 This work is
supported by  PICT 03-11609 and PIP-CONICET 6157 from Argentina,
FIS2006-08533-C03-02 and the ``Ramon y Cajal'' program (LA) from
MCEyC of Spain. CN is grateful to MEyC (Spain) for a sabbatical
stay in BIFI and Universidad de Zaragoza.

\appendix
\section{Integration of the reservoir's degrees of freedom}
In this appendix we show how to get the effective action given in eq.
(\ref{effective}). To this end we consider the path-integral
representation for the piece of the partition function that
involves the fields $\chi_r$ and $\chi_l$ to be integrated:
\begin{equation}
{\cal Z}_{res}=\int\,d \chi_r^{\dagger}\, d \chi_l^{\dagger}\, d
\chi_r \, d \chi_l \,\exp{i(S_{res}+S_{cont})}
\end{equation}
where
\begin{equation}
S_{res}= \int dx\,dt\,(\chi_r^{\dagger}D_r\,\chi_r +
\chi_l^{\dagger}D_r\,\chi_l)
\end{equation}
where $D_r$ and $D_l$ are differential operators which specific
forms depend on the dispersion relations satisfied by the
particles in the reservoir. Since the integration procedure can be
performed quite generally, we will specify the actual form for
these operators at the end of the derivation. On the other hand,
the interaction between the reservoir's modes and fermions in the
ring is given by $S_{cont}$ (see eq.(\ref{contact})). In order to
obtain an action which is quadratic in the reservoir's fields we
perform a translation in these fields of the form $\chi_{r,l}
\rightarrow  \sigma_{r,l}$ such that:
\begin{equation}
\chi_{r,l}(x,t) = \sigma_{r,l}(x,t) + a_{r,l}(x,t),
\end{equation}
(and similar transformations for $\chi_{r,l}^{\dagger}$). Choosing
the functions $a_{r,l}$ as
\begin{equation}
a_{r,l}(x,t) = -v_c\,\int
dx'\,dt'\,D_{r,l}^{-1}(x,x';t,t')\,\delta(x')\,\psi_{r,l}(x',t'),
\end{equation}
(and similar expressions for $a_{r,l}^{\dagger}$) the resulting
action is quadratic in the $\sigma_{r,l}$ fields, which are then
integrated out. Of course, there are also two terms depending
quadratically on $\psi_{r,l}$ which will also contribute to the
effective action. The general form of these action terms is:
\begin{equation}
-v_{c}^{2} \,\int
dx\,dt\,dt'\,\psi_{r,l}^{\dagger}(x,t)\,\delta(x)\,D_{r,l}^{-1}(0,0;t,t')\,\psi_{r,l}^{\dagger}(x,t).
\end{equation}
In this paper we will consider particle reservoirs such that right
and left propagators coincide for $x=x'$. We will also assume time
translation invariance such that
$D_{r,l}^{-1}(0,0;t,t')=D^{-1}(0,0;t-t')$. The reservoir's
self-energy used in the paper is defined in terms of this
propagator as $\Sigma_0(t-t')=v_c^2 \,D^{-1}(0,0;t-t')$. An
explicit form for this function is chosen in section IV C.

\end{document}